\documentclass[preprint,showpacs,preprintnumbers,amsmath,amssymb]{revtex4}

\usepackage{graphicx}
\usepackage{dcolumn}
\usepackage{bm}

\begin{document}

\title{Automating Open Fault Correction in Integrated Circuits via Field Induced Diffusion Limited Aggregation}

\author{Sanjiv Sambandan}
\email[]{ssanjiv@isu.iisc.ernet.in}
\affiliation{ Department of Instrumentation and Applied Physics, Indian Institute of Science, India}

\date{\today}

\begin{abstract}
We perform studies on electric field induced diffusion limited aggregation of conductive particles dispersed in a non conductive medium. The bridges formed across gaps between electrodes with an electric field in between due to the aggregation mechanism provides a means to automate the correction of open interconnect faults for integrated circuit application. We derive an expression for the bridging time and describe the mechanics of bridge formation with the above application in mind.
\end{abstract}

\pacs{}

\maketitle

Interconnect faults such as opens and shorts that occur during device operation are a critical element of the operational reliability of semiconductor circuits. Interconnect opens occur due to the burn out of an interconnect carrying large currents, or due to purely mechanical forces. While circuit designers generally solve these problems by replicating important hardware, there have been recent attempts to integrate a repair mechanism within the circuit by enclosing liquid metal in dielectric shells \cite{Blaiszik}. On the occurrence of an open fault due to a mechanical the dielectric shell cracks open and spills the metal onto the interconnect thereby repairing the fault.

In this letter we propose an alternative technique for open fault repair automation. We propose the integration of a dispersion of electrically conductive micro particles in an electrically insulating fluid medium within the circuit. As illustrated in Fig. 1a. the dispersion can be present within the packaging of the chip and isolated over the interconnects. On the occurrence of an open fault, for example a power line fault as shown in Fig 1b, there would appear an electric field in the newly formed gap depending on the potential of the line. The conductive particles polarize in the presence of the field begin to drift and aggregate due to the dielectrophoretic forces and dipole interaction \cite{Wen}-\cite{dep7}. Through a drift-diffusion process, the conductive particles eventually bridge the gap between the two electrodes thereby repairing the open fault. Thus, the repair mechanism is triggered by the occurrence of an open fault. In this letter we study this repair mechanism and characterize the bridging time.

The drift component of this drift-diffusion process is the resultant of three forces. The first is the dielectrophoretic force, $F_{e}$ on the conductive particles due to the applied electric field which draws the conductive particles towards regions of high electric field. Assuming the conductive particles are spherical with radius $R$, if the open fault has an electric field $\xi$ across it, 
\begin{equation}
F_{e}=2\pi R^{3} \epsilon_{m} \mbox{Re}[\beta] \nabla |\xi|^{2} 
\end{equation}
Here $\beta=\frac{\epsilon_{p}-\epsilon_{m}- j\frac{\sigma_{p}-\sigma_{m}}{\omega}}{\epsilon_{p}+2\epsilon_{m}- j\frac{\sigma_{p}+2\sigma_{m}}{\omega}}$ is the Clausius-Mossotti factor \cite{dep1}-\cite{dep7}. If the particle and medium have a large difference in conductivity with $\sigma_{p}>>\sigma_{m}$ $\mbox{Re}\beta \approx 1$, and the particles move towards the region of high electric field. Due to the finite dimensions of the interconnect, this electric field is non-uniform and $F_{e} \neq 0$. The second force resulting in the drift of the particles is the dipole-dipole interaction between the individual particle dipoles, $F_{d}$ resulting in attraction and aggregation \cite{Wen}, \cite{ER}, \cite{dip}. If one of the polarized particles is at the origin, and the other at a distance $r>>R$ from it, with the line joining their centers making an angle $\theta$ with the direction of the electric field, the outward radial component of $F_{d}$ is 
\begin{equation}
F_{d}= 12 \pi \epsilon_{m} \beta^{2} \xi^{2} \frac{R^{6}}{r^{4}}(1-3\mbox{cos}^{2}(\theta))
\end{equation}
Thus the particles lying more parallel to the electric field lines ($\mbox{cos}(\theta)>1/\sqrt{3}$) experience an attractive force and aggregate. The third force is the viscous drag, $F_{v}$ experienced by the particles in the fluid medium and is given by $F_{v}= \mu u$, where $\mu=6\pi \eta R $, with $\eta$ being the viscosity of the fluid medium, and $u=dr/dt$ is the velocity of the particle in the fluid medium. The diffusion coefficient, $D$ of the drift-diffusion mechanism is defined by the Einstein relation $D=kT/\mu$. In summary, $F_{e}$ results in gathering all the conductive particles in regions where the electric field gradient is maximum. $F_{d}$ then chains the particles together till the bridge is formed and the field is dissipated. $F_{v}$ opposes all the motion of the particles. $F_{d}$ is particularly strong at short distances, and $F_{e}$ becomes comparable to $F_{d}$ when $r \geq (12 \xi^{2} R^{3} / \nabla |\xi|^{2})^{1/4}$ (for $\theta \approx 0$).

\begin{figure}
 \includegraphics[]{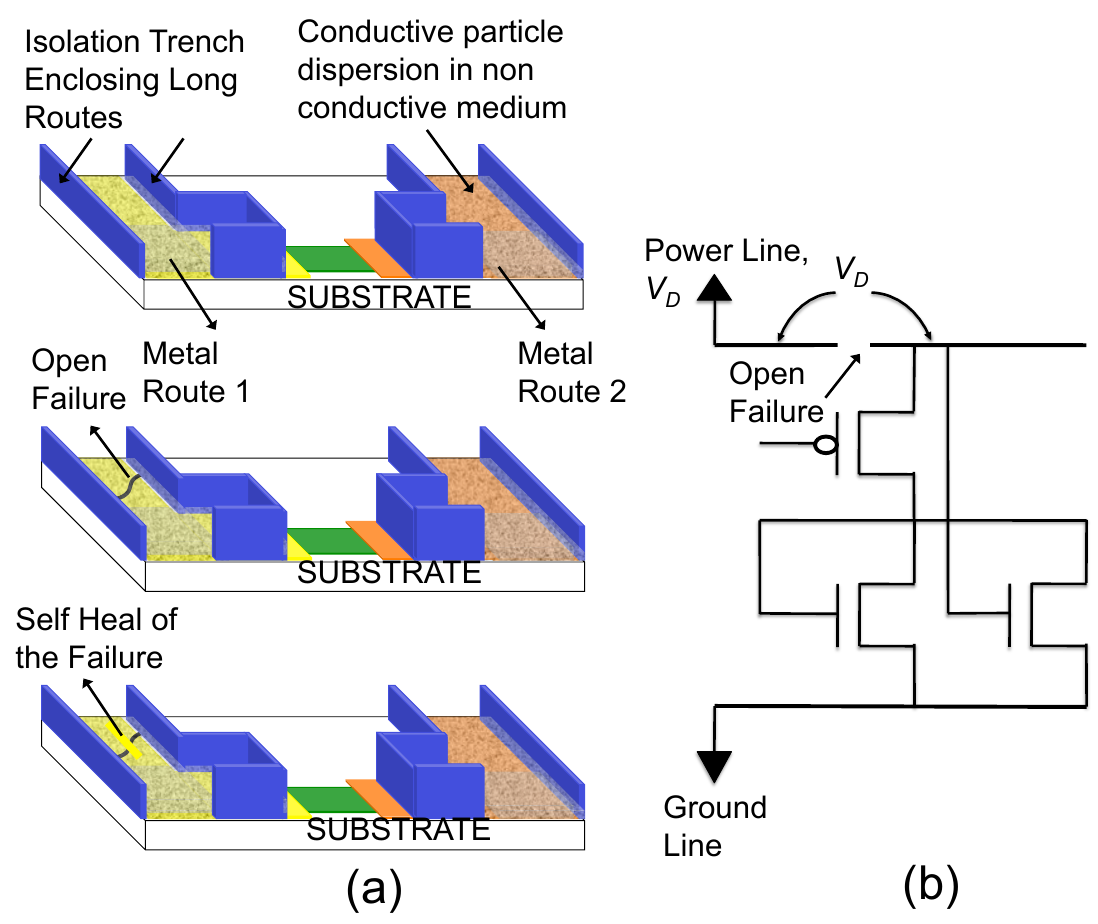}
 \caption{(a) Illustration of the integration of the open fault repair mechanism within the circuit. A dispersion of conductive particles in a non conductive medium is placed over the interconnects with isolation. (b) An open fault leading in a potential difference across the gap automatically triggers the aggregation and eventual bridging of the interconnect.}
\end{figure}

There are several tradeoffs in the selection of the medium. A medium with high permittivity adds capacitive components along the interconnect and is undesirable. The conductivity of the medium must also not be high since it reduces $F_{e}$ and therefore increases the heal time of an open fault. Moreover, it can lead to unwanted shorts and parasitic resistive loading. In a medium with low viscosity the dispersed particles tend to aggregate into clusters more easily since the viscous drag becomes small. This is undesirable since we would like the dispersion to have a uniform density through the lifetime of the circuit. Thus, the preferable medium must have moderately high viscosity, low permittivity and low conductivity. We demonstrate the concept using a dispersion of metallic carbon nanotubes (CNTs) of approximately 1 micron length and 20-50nm diameter, and zinc microparticles of approximate diameter 10-20 micron, in silicone oil. The oil used had a kinematic viscosity of 300cS and density of 0.97gm/ml. The dispersion was achieved by sonicating mechanically powdered particles in hot silicone oil for about 1 hour. Other techniques used were to first sonicate the particles in iso-propyl alchohol followed by a drying of the dispersion. The dried out particles were then dispersed in silicone oil.

\begin{figure}
 \includegraphics[]{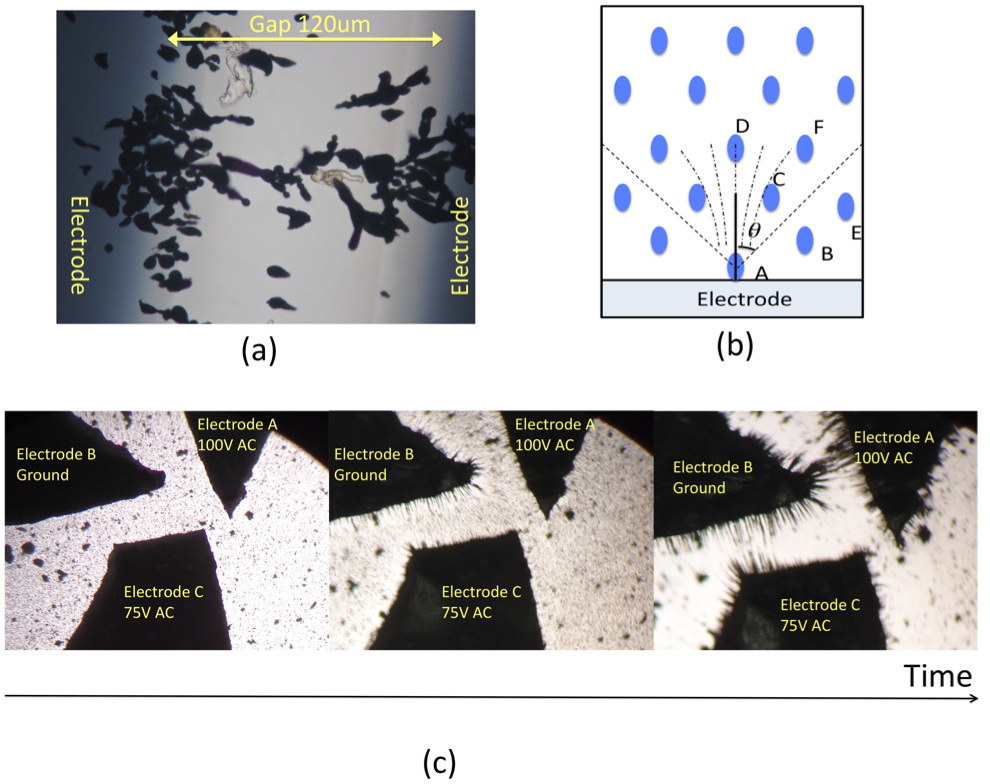}
 \caption{(a) Typical bridges between the electrodes formed by Zn microparticles, (b) Illustration of the mechanics of bridge formation. Particle A forms a seed for cluster growth at the electrode. All polarized particles within the subtended angle of $2\theta=2\mbox{cos}^{-1}(1/\sqrt{3})$, are attracted due to dipole interactions. (c) Bridging between three electrodes subject to electric fields oscillating at 50Hz.}
\end{figure}

\begin{figure}
 \includegraphics[width=4in]{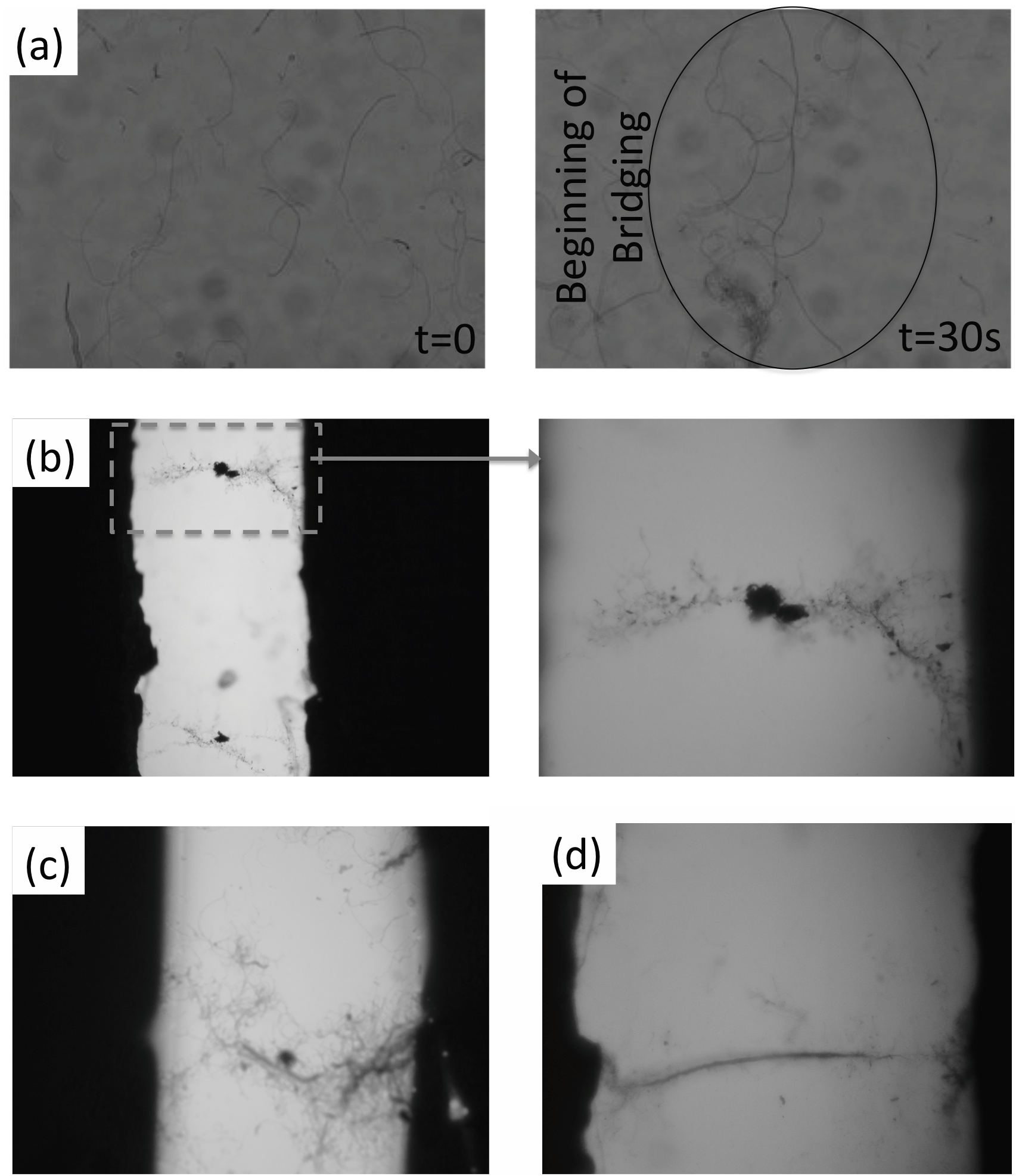}
 \caption{(a) Beginning of the bridging process with CNTs. The figure on the left is captured just before the application of the electric field. The figure on the right is captured about 60s after the application of the field (0.25V/micron), (b)-(d) Typical bridges formed with CNTs.}
\end{figure}

Fig. 2 illustrates typical bridge formations between two electrodes for a dispersion of 65mg/ml of zinc microparticles in silicone oil. In Fig. 2a, the voltage across the electrodes is 30V dc and the gap is 120 micron. The typical growth process is expected to be as described by Fig.2b. Upon the application of the electric field, the nearest particles to the electrodes are polarized and stick to the electrode (for example A in Fig.2b) forming a seed for growth. All particle dipoles within the angles defined by $2\theta=2cos^{-1}(1/\sqrt{3})$, (for example C) are attracted to particle A and begin to drift towards the electrodes while undergoing diffusive motion. Particles that are outside this zone (for example B and E) are pushed away and form seeds themselves. The non uniform electric field caused by A (shown by dotted curving lines) results in further attraction of polarized particles due to $F_{e}$. As the clusters grow, the field distribution in the gap becomes more non uniform due to the sharp nodes created by the clusters. With time, clusters are formed in the gap far from the electrodes and fractal like growth occurs at the electrodes when these clusters attach themselves to the seed. In Fig. 2c, three electrodes are at 50Hz ac potential of 100V rms, 75V rms and ground. The separation between the electrodes is about 1000 um. The particles in the dispersion first seek and bridge the electrodes with the maximum field in between the gap. Thus, there is a hidden search mechanism in the fault repair mechanism where, the opens that have the maximum field are bridged first. Fig. 3 illustrates the bridge formation between two electrodes with a dispersion of 0.6mg/ml of metallic CNTs in silicone oil. Fig 3a shows the beginnings of the bridging process where the nanotube clusters begins to aggregate and link up on the application of a field. Fig. 3(b-d) show typical CNT bridges formed between the electrodes.

\begin{figure}
 \includegraphics[]{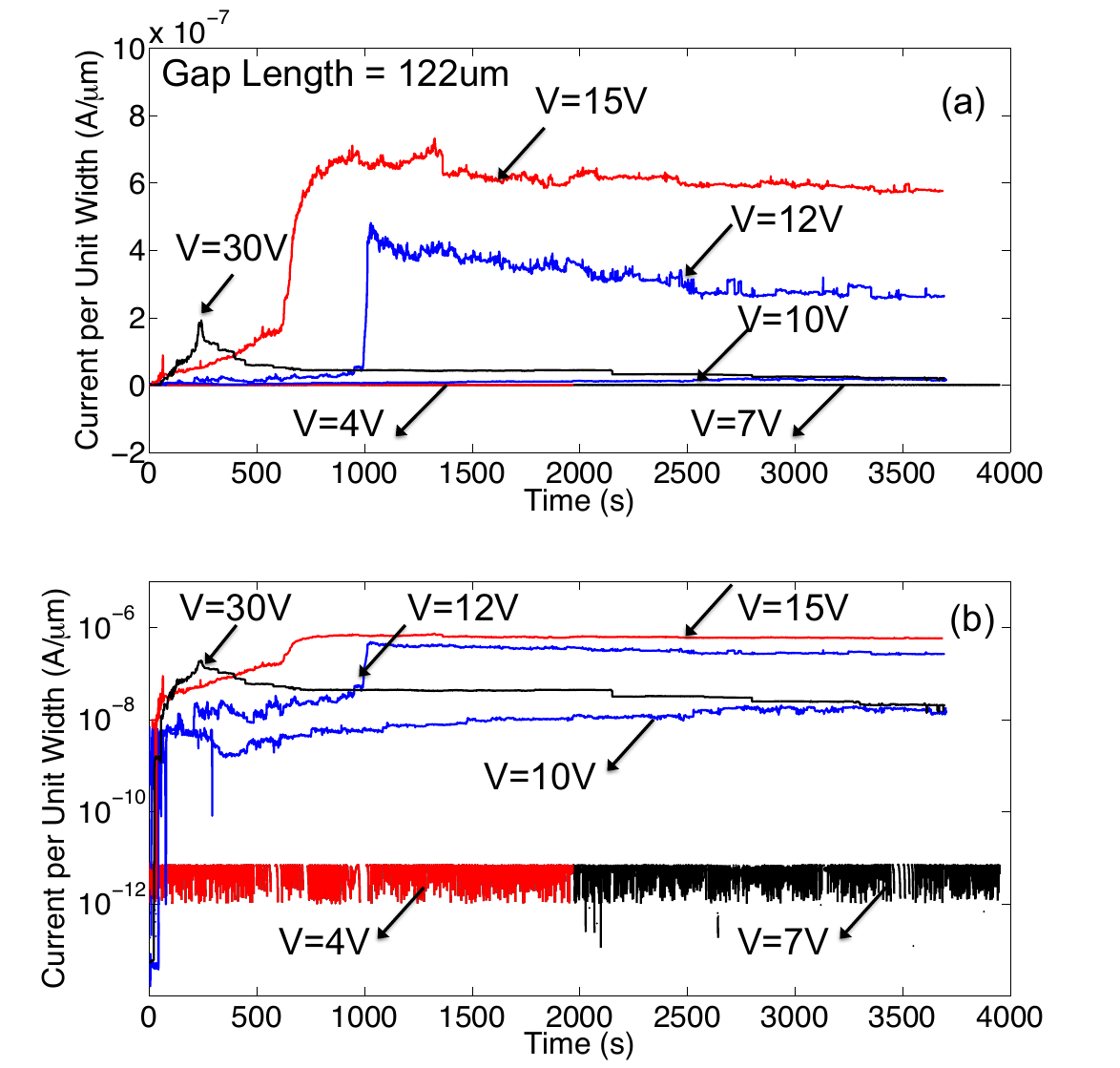}
 \caption{Current density through the gap as a function of time for different electric fields using CNT dispersion. The gap length is fixed at 120 micron. (a) Current density on a linear scale, (b) current density on a log scale.}
\end{figure}

\begin{figure}
 \includegraphics[]{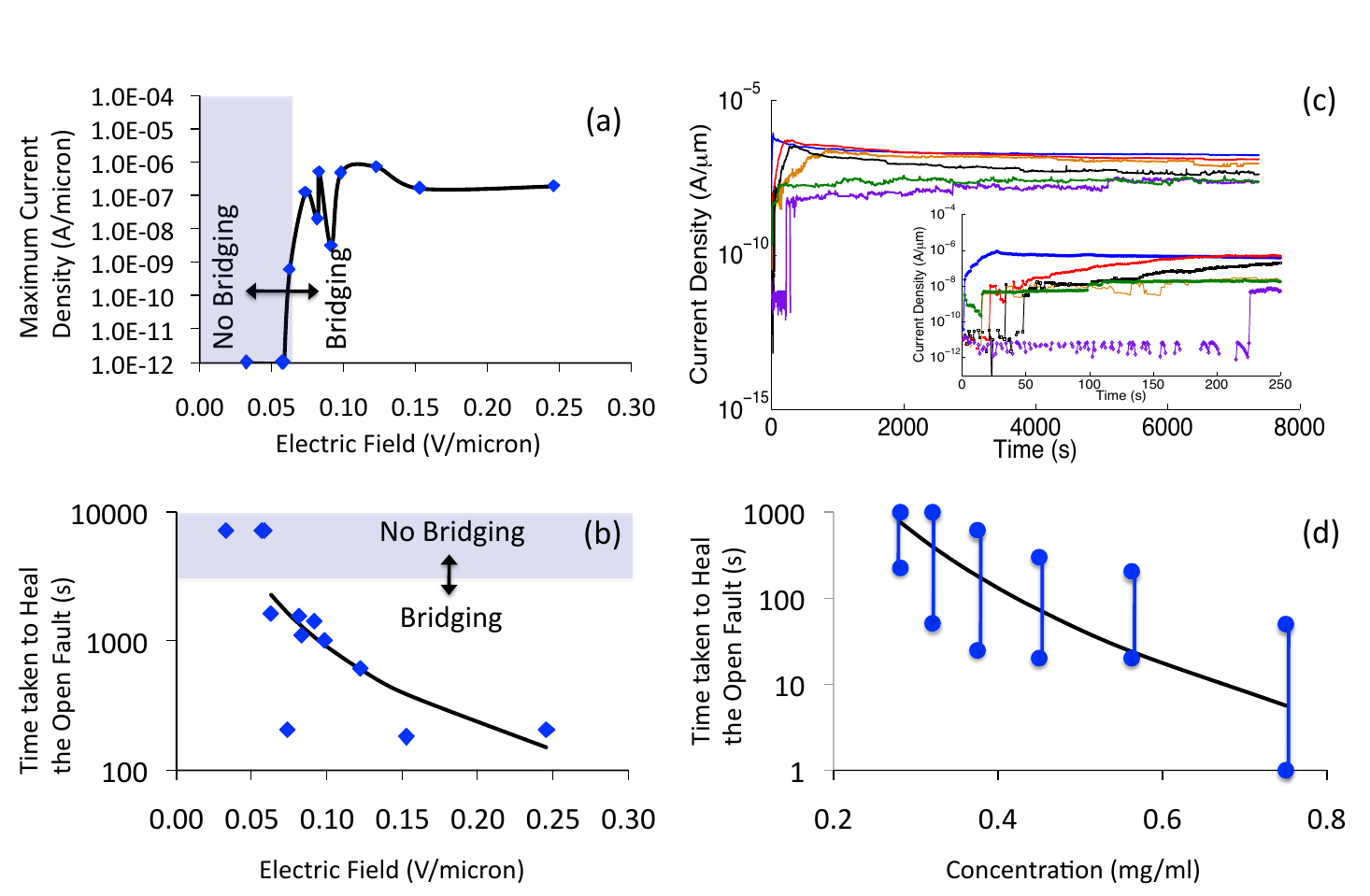}
\caption{(a) Maximum current density of the bridges as a function of the electric field for various applied voltages and various gap lengths. (b) Bridging time as a function of electric field. Incomplete bridges are labeled at 7200s, which was the duration of the experiment. The bold line represents the model for bridging time. (c) Current density as a function of time, for bridging at different concentrations. The electric field is kept constant at $\approx 0.25$V/micron. The inset shows a zoom in at the low time scales. Bridging occurs rapidly, and then slowly becomes stronger till the maximum current density is achieved. (d) Summary of Fig. 5c. Plot of the bridging time as a function of concentration. The dark line represents the model. At each concentration we have a minimum bridging time point - when the bridge is complete and enables current flow - and a maximum bridging time point - representing the time taken to reach maximum current flow. }
\end{figure}

Applying different dc voltage across electrodes separated by a 120 micron gap, we measure the current through the gap as a function of time. The electrode width is kept large (about 1e4um) so as to average out the effects of non uniformity in the dispersion. Fig. 4a and 4b show the typical dynamics of current with the current data plotted on a linear and log scale, respectively. For a fixed dispersion concentration, the key parameter influencing the successful formation of the bridge within the observed time is the electric field. It was seen that the bridging time decreases with the application of higher fields, where the bridging time is defined as the time taken for the current to rise to its maximum value. For very high fields (for example at 30V), it was consistently observed that the bridges were unstable. Under these high fields the current first rose to the maximum value before dropping off. While we attribute this behavior to thermal breakdown of bridges, more studies are needed to justify this explanation. The bridges are most stable for moderately high electric fields. Fig 5a summarizes the maximum current density capacity of the bridges as a function of the field strength for electrode gap lengths and gap voltages. The threshold electric field for this dispersion concentration was observed to be about 0.075V/micron. Fig. 5b plots the typical bridging time for various electric fields. Fig. 5c and 5d plot the dynamics of maximum current density and bridging time for different concentrations of the dispersion with the electric field held constant at $0.25$V/micron. The inset of Fig. 5c shows a zoom in at the low time scales. Bridging occurs rapidly first forming a path, and then slowly becomes stronger till the maximum current density is achieved. Thus the bridging time is represented by two time points, a minimum bridging time point - when the bridge is complete and enables current flow - and a maximum bridging time point - representing the time taken to reach maximum current flow as shown in Fig. 5d. 

Revisiting the mechanics described in Fig. 2b we can build an approximate model for the bridging time. Consider the dispersion to be uniform with an average separating distance between the particles being $r_{0}$. After a particle is attached to one of the electrodes, it forms a seed for growth. All polarized particles (say $c$ particles) within the subtended angle are attracted to the seed particle and travel the distance $r_{0}-R$ to attach to the seed. Assuming that the dipole interactions are the dominant force component, the effective velocity of the particles is $u_{eff}=F_{d}/\mu$, and the time taken for the $c$ particles to reach the seed is $\approx (r_{0}-R)/u_{eff}$. After these $c$ particles are attached to the seed, the seed cluster advances through the gap by $\approx c^{\frac{1}{\gamma}}R$. If the growth was two dimensional, $\gamma=2$, however,  it has been shown that this cluster growth has fractional dimensions ($1<\gamma<2$) with the dimension scaling with the volume fraction of the dispersion \cite{Wen}. In this time, $c$ polarized particles in the medium aggregate into clusters but generally dont move because they are balanced by the forces pulling it towards the seed cluster at the electrode and those applied by their neigbors in the medium. The average separation of particles in the medium is now $cr_{0}$. These clusters now have to travel a distance $\approx (cr_{0}-c^{\frac{1}{\gamma}}R)$ in order to attach themselves to the seed cluster at the electrode. This process continues and the time taken for the $j^{\mbox{th}}$ cluster to reach the seed cluster at the electrode is $\tau_{j}=(c^{j-1}r_{0}-c^{\frac{j-1}{\gamma}}R)/u_{eff}$. This process occurs from both electrodes simultaneously and the bridging is completed when the bridge from one electrode reaches the halfway point covering a distance  $d/2$ where $d$ is the gap length. When the bridging is completed,$(c^{\frac{j}{\gamma}}-1)R/(c-1) = d/2$ and the total bridging time, $\tau_{b}$ is given by
\begin{eqnarray}
\tau_{b} &\approx& \left(\frac{d}{2R}\right)^{\gamma}(c-1)^{\gamma-1}\frac{r_{0}}{u_{eff}} \\ \nonumber
&\approx& \left(\frac{d}{2R}\right)^{\gamma}(c-1)^{\gamma-1}\left(\frac{r_{0}}{R}\right)^{5}\frac{1}{\xi^{2}}\frac{\eta}{2\epsilon_{m}} 
\end{eqnarray}
We extract $\gamma$ by relating finding the effective area covered, $A$ by the particles in a square of length $L$ and noting, $L^{\gamma}=A$. Using $c=2$, $d/R=120$, $r_{0}/R \approx 3$ at a concetration of 0.6mg/ml, $\epsilon_{m}=22.12e-12$F/m, $\beta \approx 1$, $\eta=300e-6$Ns/m$^{2}$, and $\gamma \approx 1.8$, the model (solid line) is plotted against experimental results in Fig. 5b and Fig. 5d.

In conclusion, we present a possible solution of integrating dispersions of conductive particles in semiconductor circuit packages to automate the correction of line open faults. The packaging and actual implementation would have to borrow ideas from the techniques used in the fabrication of micro fluidic lab-on-chip systems.

We would like to thank Dr. A. Misra for providing us with the carbon nanotubes.

\end{document}